\documentclass[12pt,preprint]{aastex}

\shorttitle{Formation Channel to PSR J1903$+$0327} \shortauthors{Liu
et al.}

\begin{document}

\title{A Fallback Disk Accretion-Involved Formation Channel to
    PSR J1903$+$0327}


\author{Xi-Wei Liu\altaffilmark{1} and Xiang-Dong Li\altaffilmark{2}}

\altaffiltext{1}{Institute of Astrophysics, Huazhong Normal
University, Wuhan 430079, China} \altaffiltext{2}{Department of
Astronomy, Nanjing University, Nanjing 210093, China}

\email{liuxw@phy.ccnu.edu.cn; lixd@nju.edu.cn}

\begin{abstract}
The discovery of the eccentric binary and millisecond pulsar PSR
J1903$+$03273 has raised interesting questions about the formation
mechanisms of this peculiar system. Here we present a born-fast
scenario for PSR J1903$+$03273. We assume that during the supernova
(SN) explosion that produced the pulsar, a fallback disk was formed
around and accreted onto the newborn neutron star. Mass accretion
could accelerate the neutron star's spin to milliseconds, and
decrease its magnetic field to $\sim 10^8-10^9$ G, provided that
there was sufficient mass ($\sim 0.1 M_{\sun}$) in the fallback
disk. The neutron star became a millisecond pulsar after mass
accretion terminated. In the meanwhile the binary orbit has kept to
be eccentric (due to the SN explosion) for $\sim 10^{9}$ yr. We have
performed population synthesis calculations of the evolutions of
neutron stars with a fallback disk, and found that there might be
tens to hundreds of PSR J1903$+$03273-like systems in the Galaxy.
This scenario also suggests that some fraction of isolated
millisecond pulsars in the Galactic disk could be formed through the
same channel.
\end{abstract}





\section{Introduction}

There are now around 100 binary and millisecond pulsars (BMPSRs)
known in the Galaxy. Most of them are characterized by short spin
periods ($P\la 15$ ms), weak magnetic fields ($B\sim 10^8-10^9$ G),
and circular orbits with companions of mass $\sim 0.15 M_{\sun}-0.45
M_{\sun}$. In the recycling scenario, these systems are thought to
be the descendants of low-mass X-ray binaries (Bhattacharya \& van
den Heuvel 1991; Tauris \& van den Heuvel 2006). The evolutionary
path is briefly described as follows. A high-field ($B\sim
10^{12}-10^{13}$ G), rapidly rotating neutron star (NS) is born in a
binary with a low-mass ($\sim 1 M_{\sun}$) main-sequence (MS)
companion star. During the supernova (SN) that produces the NS, mass
loss and a kick imparted on the NS cause the orbit to be eccentric.
The NS spins down as a radio pulsar for $\sim 10^6-10^7$ yr until
passing by the so-called ``death line" in the magnetic field - spin
period ($B-P$) diagram \citep{ruderman75}. When the companion
evolves to overflow its Roche lobe, mass transfer occurs by way of
an accretion disk, and tidal friction serves to circularize the
orbit. Mass accretion onto the NS gives rise to X-ray emission,
induces magnetic field decay (the mechanisms for the field decay
induced by accretion are, however, not well understood), and
accelerates the NS up to short (millisecond) period. When the
companion loses almost all of its envelope and mass transfer ceases,
the endpoint of the evolution is a circular binary containing an NS
visible as a low-field, millisecond radio pulsar, and a He or CO
white dwarf (WD), the remaining core of the companion. A
fast-growing population of eccentric ($e>0.1$) BMPSRs have been
recently revealed in globular clusters (GCs) \citep{freire07}. These
high eccentricities are likely to be attributed to the dynamical
interactions in the central regions of GCs.

Most recently \citet{champion08} reported the discovery of an
eccentric BMPSR PSR J1903$+$0327 in the Galactic plane.  With a spin
period of 2.15\,ms, the pulsar lies in an eccentric ($e=0.44$),
95-day orbit around a $\sim 1 M_{\sun}$ companion. The mass of the
pulsar was estimated to be $1.74\pm0.04\,M_{\sun}$, about $30\%$
larger than that of other binary NSs in the Galactic disk. Infrared
observations identified a possible MS companion star. Preferred
formation scenarios may include \citep[cf.][]{heuvel08,champion08}:
(1) the pulsar was recycled in a GC, then the original donor star
was replaced by the present MS companion via one or more exchange
interactions, and the binary was displaced from the GC due to either
ejection  or the disruption of the GC; (2) the pulsar is part of a
primordial hierarchical triple system, recycled by accretion from
the progenitor of a massive ($\sim 0.9-1.1\,M_{\odot}$) WD which is
seen in the timing measurement, while the detected infrared
counterpart, the third star, is in a much wider and highly inclined
orbit around the inner binary.  The eccentricity of the inner binary
is caused by the perturbation of the outer MS star; (3) in an
alternative triple-star model, the WD in the inner binary system was
evaporated by the pulsar's energy flux, or coalesced with it, such
that the present binary remained.

% 4) The companion
%star is a pulsar, though its radio pulsation is not detected due
%to either unfavorably beaming away from our line of sight or
%rotating too slowly to be observed.

Since there is no strong evidence for the GC or triple star origin
of PSR J1903$+$0327, in this paper, we seek an alternative,
born-fast scenario for PSR J1903$+$0327 which was already mentioned
by \citet{champion08}. These authors argued that the pulsar was not
likely to be formed spinning rapidly at the time of SN with a small
magnetic field, due to the following reasons: (1) there are no
pulsars like PSR J1903$+$0327 in any of the $>50$ young supernova
remnants in which an NS has been inferred or detected directly; (2)
the 18 isolated MPSRs detected in the Galactic disk have spin
distributions, space velocities, and energetics indistinguishable
from those of recycled BMPRSs and their space velocities and scale
heights do not match those of non-recycled pulsars; (3) magnetic
fields in young pulsars is unlikely to be less than $10^{10}$ G.
However, as we shown below, if experienced accretion from a SN
fallback disk, the newborn pulsar could be accelerated to a period
of several milliseconds, together with the magnetic field decayed to
$\sim 10^{8}$\,G from an initial value, say $\sim 10^{12}$ G. When
the disk becomes neutral after $\sim 10^{3}-10^{4}$\,yr
\citep{menou01} and accretion ceases, the NS becomes an MPSR. The
orbit, if not being disrupted by the SN explosion, will remain
eccentric for $> 10^{10}$\,yr before it is circularized
\citep[cf.][]{hurley02}. The binary consists of a recycled pulsar
and a MS companion in an eccentric orbit, just like PSR
J1903$+$0327. Hence this scenario could produce a population of
eccentric BMPSRs as well as isolated MPSRs.

%In the next section we examine whether the fallback model can
%reproduce the properties of PSR J1903+0327.  In section 3 we
%calculate the statistical properties of this kind of FDAI MPSRs.
%Conclusions and discussions are given in the last section.

\section{Spin Evolution of An NS with A Fallback Disk}

%% In a manner similar to \objectname authors can provide links to dataset
%% hosted at participating data centers via the \dataset{} command.  The
%% second curly bracket argument is printed in the text while the first
%% parentheses argument serves as the valid data set identifier.  Large
%% lists of data set are best provided in a table (see Table 3 for an example).
%% Valid data set identifiers should be obtained from the data center that
%% is currently hosting the data.
%%
%% Note that AASTeX interprets everything between the curly braces in the
%% macro as regular text, so any special characters, e.g. "#" or "_," must be
%% preceded by a backslash. Otherwise, you will get a LaTeX error when you
%% compile your manuscript.  Special characters do not
%% need to be escaped in the optional, square-bracket argument.

Following the formation of an NS through the core collapse of its
progenitor, a small amount of mass could fall back onto the compact
object \citep{colgate71,chevalier89,lin91}.  Some of the fallback
material may carry sufficient angular momentum and to form an
accretion disk of mass M$_{d}$ around the NS. The initial transient
phase lasts for a local viscous timescale $t_{0}\approx
6.6\times10^{-5}(T_{c,6})^{-1}R_{d,8}^{1/2}(t_{0})$\,yr, on which a
thin disk forms \citep{cannizzo90,menou01}. Here $T_{c,6}$ is the
typical temperature in the outermost disk annulus during this early
phase, in units of $10^{6}$\,K, and $R_{d,8}(t_{0})$ is the initial
disk radius in units of $10^{8}$\,cm.  We set $T_{c,6}=1$ throughout
this paper.  The subsequent evolution of the disk obeys the
self-similar solution \citep{cannizzo90}, i.e.,
\begin{equation}
\dot{M}_{d}(t) =\left\{
\begin{array}{ll}
\dot{M}_{d}(t_{0}), & 0<t<t_{0}, \nonumber \\
 \dot{M}_{d}(t_{0})\left(\frac{t}{t_{0}}\right)^{-p}, &
 t\geq t_{0},
 \end{array}\right.
\end{equation}
where the power index $p=1.25$ \citep{fran02},
$\dot{M}_{d}(t_{0})$ is a constant, which we normalize to the
total mass of the disk $M_{d}=\int^{\infty}_{0}\dot{M}_{d}dt$, by
$M_{d}/(5t_{0})$ \citep{chatterjee00a}.

The subsequent evolution of the NS can be divided into three phases:

1. The {\it accretor} phase - When the magnetosphere radius
$R_{m}\simeq 1.6\times10^{8}B_{12}^{4/7}\dot{M}_{18}^{-2/7}$\,cm is
less than the corotation radius defined by
$R_{c}=(GM/\Omega)^{1/3}$, the fallback matter is allowed to be
accreted onto the surface of the NS. Here $B_{12}=B/10^{12}$ G,
$\dot{M}_{18}=\dot{M}/10^{18}$ gs$^{-1}$, and $\Omega$ is the
angular velocity of the NS.  To evaluate the accretion torque, we
divide this phase further into two sub-phases: a) If $R_{m}\leq$ the
NS radius $R_{NS}$, the torque exerted on the star is assumed to be
$\dot{J}=\dot{M}R_{NS}^{2}\Omega_{K}(R_{NS})$, where $\Omega_{K}(R)$
is the Keplerian velocity at $R$;  b) if $R_{NS}<R_{m}\leq R_{c}$,
the accretion torque is given by
$\dot{J}=2\dot{M}R_{m}^{2}\Omega_{K}(R_{m})[1-\Omega/\Omega_{K}(R_{m})]$.

The accretion rate $\dot{M}$ of the NS is assumed to be limited by
the Eddington accretion rate $\dot{M}_{E}\simeq
1\times10^{18}$\,g\,s$^{-1}$ for a $1.4\,M_{\sun}$ NS. However,
during the early phase of accretion, the accretion rate may be so
high that the radiation is  trapped in the flow and neutrino losses
are important close to the NS surface \citep{colgate71,chevalier89}.
The gravitational energy generated by infalling matter is mainly
released by neutrino losses.  Hence the Eddington limit does not
work in this phase. According to \citet{chevalier89}, when the
accretion rate drops to about $3\times10^{-4}\,M_{\odot}$ yr$^{-1}$,
the reverse shock reaches the radiation trapping radius so that the
photons can begin to diffuse out from the shocked envelope.  A
luminosity of Eddington limit is expected now, and this luminosity
can reduce or even reverse the inflow outside the shocked envelope.
Hence we adopt the Eddington limited accretion only when
$\dot{M}<3\times10^{-4}\,M_{\odot}$\,yr$^{-1}$. Otherwise we assume
all the mass is accreted by the NS.

Along with mass accretion, we assume that the magnetic field
evolution follows the relation,  $B=B_{0}(1+\Delta M/M_{0})$, where
$B_0$ is the initial magnetic field strength, and $M_{0}$ is a
parameter that determines the rate of decay with its typical value
of $\sim 10^{-5}-10^{-4}M_{\odot}$
\citep{taam86,shibazaki89,romani90}.

2. The {\it propeller} phase - This phase begins when $\dot{M}$
decreases so that $R_{c}<R_{m}$. The infalling matter is assumed to
be accelerated outward owing to the centrifugal barrier, taking away
the angular momentum of the NS \citep{illarinov75}.  In this phase
we use the same formula as in the {\it accretor} phase b) to
estimate the propeller spin-down torque.

3. The {\it radio pulsar} phase - As $\dot{M}$ decreases further,
$R_{m}$ exceeds the light cylinder radius $R_{lc}=c/\Omega$, or
the disk becomes neutral,  the accretion ceases and the NS becomes
a radio pulsar.  The neutral timescale of a fallback disc is
$t_{n}=[R_{d,10}^{3}(t_{0})\times10^{16}\mathrm{g\,s}^{-1}/\dot{M}_{d}(t_{0})]^{-1/2.75}t_{0}$
\citep{menou01}, where $R_{d,10}$ is the disk radius in units of
$10^{10}$\,cm. In this work, when either $R_{m}>R_{lc}$ or
$t>t_{n}$, we assume that the radio pulsar phase turns on and the
subsequent NS spin evolution follows the magnetic dipole radiation
prescription, until the NS crosses the "death line", i.e.
$B_{12}/P^{2}<0.17$.

In Fig.~1 we show several examples of the calculated evolutions of
the NS spin and magnetic field under typical conditions according
the scheme described above.  The model parameters are listed in
Table 1. For the spin evolution, the thick solid, solid, and dotted
lines represent the {\it accretor}, {\it propeller}, and {\it radio
pulsar} phases, respectively.  In case 1 we adopt the initial
parameters as $M_{d}=0.28\,M_{\odot}$, $P_{0}=14$\,ms,
$B_{0}=3\times10^{12}$\,G, $M_{0}=10^{-5}\,M_{\odot}$, and
$R_{d}=10^{9}$\,cm. The NS is found to be spun-up to a period of
2.88\,ms in 72\,yr, when the disk becomes neutral\footnote{The
neutral time is given by
$t_{n}\propto[R_{d}^{-3}\dot{M}_{d}(t_{0})]^{1/2.75}\times
R_{d}^{0.5}$ \citep{menou01}. The value $t_{n}=72$\,yr used here is
much less than that ($\sim 10^{3}\sim10^{4}$\,yr) in
\citet{menou01}.  This is because (1) we calculated
$\dot{M}_{d}(t_{0})$ by normalizing it to the total disk mass, while
\citet{menou01} estimated it by assuming
$\dot{M}_{d}(t_{0})=M_{d}/t_{0}$, which gives a larger value of
$\dot{M}_{d}(t_{0})$ than ours; (2) the adopted value of
$R_{d}=10^9$ cm is larger than the value in \citet{menou01}.}.
 In the meanwhile, the magnetic
field decreases dramatically to $4.2\times10^{8}$\,G. This is a
strongly spun-up pulsar, demonstrating a possible way for the
formation of PSR J1903+0327. In case 2 we lower the initial disk
mass $M_d$ to $0.02M_{\odot}$, while keeping all the other
parameters same as in case 1.  This leads to a lower
$\dot{M}_{d}(t_{0})$, so that the spinning-up efficiency also
becomes lower.  The NS enters the {\it radio pulsar} phase with a
relatively longer period of 11.26\,ms and a higher magnetic field of
$6.8\times10^{9}$\,G. In case 3 we lower the initial disk radius
$R_d$ by a factor of 10 compared with in case 1.  This gives a lower
value of $t_{0}$ and a higher value of $\dot{M}_{d}(t_{0})$, and the
NS is spun-up to a period of 3.63\,ms. In case 4 we choose a slower
decay of the magnetic field than in case 1 by setting a higher value
of $M_{0}=10^{-4}\,M_{\sun}$.  It is shown that the period evolution
remains almost unchanged, but the magnetic field decays to
$4.2\times10^{9}$\,G, about 10 times of that in case 1. A similar
feature can be seen in case 5, if we increase the initial magnetic
field to be $B=1\times10^{13}$\,G. Finally in case 6 $M_{d}$ is
further decreased to be $0.01\,M_{\odot}$, now the NS is spun-up to
a period of 11.82\,ms during the first 39 yr and then spun-down
during the {\it propeller} phase to a period of 11.92\,ms during the
subsequent 88\,yr. After that time the disk becomes neutral and the
{\it radio pulsar} phase begins.  In the meanwhile the magnetic
field decays to $6.6\times10^{10}$\,G. This is a mildly spun-up
pulsar.

%% In this section, we use  the \subsection command to set off
%% a subsection.  \footnote is used to insert a footnote to the text.


%% This section also includes several of the displayed math environments
%% mentioned in the Author Guide.

\section{Population synthesis}

Based on the analysis in the above section, now we try to estimate
the statistical properties (e.g. the birth rate, total number,
orbital period and eccentricity distributions) of RSR
J1903+0327-like pulsars formed through the fallback disk accretion
involved (FDAI) channel in the Galaxy, by using an evolutionary
population synthesis (EPS) method based on the rapid binary stellar
evolution (BSE) code developed by \citet{hurley00} and
\citet{hurley02}. This code incorporates the evolution of single
stars with binary-star interactions, such as mass transfer, mass
accretion, common-envelope (CE) evolution, SN kick, tidal friction,
and angular momentum loss mechanisms (e.g. magnetic braking and
gravitational radiation).

We assume all stars are born in binary systems.  The initial mass
function (IMF) of \citet{kroupa93} is adopted for the primary's mass
($M_{1}$) distribution.  For the secondary stars ($M_{2}$) we assume
a uniform distribution of the mass ratio $M_{2}/M_{1}$ between 0 and
1.  A uniform distribution of $\ln a$ is adopted for the binary
separation $a$.  The star formation rate parameter is taken to be
$S=7.6085$\,yr$^{-1}$, corresponding to a rate of
$\sim 0.02$\,yr$^{-1}$ for core-collapse SNe in our Galaxy, assume
all the stars with masses $>8\,M_{\odot}$ die through SN
explosions.

It is known that a velocity kick is imparted to newborn NSs during
core-collapse SNe, and a Maxwellian SN kick distribution with
dispersion $V_{\sigma}\sim 200-400$ km\,s$^{-1}$ is usually adopted
\citep[e.g.][]{lyne94,hobbs05}. However, there is mounting evidence
that some NSs may have received relatively small velocity kicks, of
order less than 50\,km\,s$^{-1}$ \citep{pfahl02a,pfahl02b,pod04}. A
probable mechanism that causes small kicks is the electron capture
(EC) SNe \citep{miyaji80,nomoto84}. The explosion energies of such
SNe are significantly lower than those of normal SNe
\citep{dessart06,kitaura06,jiang07}, so that the nascent NSs receive
a small kick velocity.  It has been suggested that the initial
masses of stars that explode via the EC SN mechanism may range in
$\sim 6-12\,M_{\odot}$ \citep{pod04,kitaura06,poe08}. Alternatively,
core-collapse SN would be sub-energetic if the initial mass of its
progenitor is greater than $\sim 18\,M_{\odot}$. Simulations for a
range of progenitor masses and different input physics suggest that
the explosion energy decreases with increasing progenitor mass,
whereas the binding energy of the star increases with increasing
mass.  This leads to a transition mass range of $\sim
18-25\,M_{\odot}$ above which the SN explosion is so
energy-deficient that fallback accretion would be important
\citep{fryer99}.  For the aforementioned reasons, we assume a
Maxwellian kick distribution with $V_{\sigma}=20$\,km\,s$^{-1}$ for
the progenitor stars with initial masses of $8\leq
M\leq12\,M_{\odot}$ and $M\geq 18\,M_{\odot}$, otherwise we take
$V_{\sigma}=265$ km\,s$^{-1}$.

The fallback disk mass $M_{d}$ is poorly known, but various
arguments suggest that $M_{d}\la 0.1\,M_{\odot}$ may not be
unreasonable \citep{chevalier89,lin91}.  There should exist a
relation between the disk mass and the properties of the
progenitors. Unfortunately, there are big uncertainties in some key
factors in the SN models, including the neutrino energy and
transport, equation of state of gas at nuclear densities,
convection, rotation, neutrino emission asymmetry, etc
\citep{monchmeyer91,yamada95,fryer99}, each of which could strongly
affect the SN energy and hence the mass of fallback.  So we do not
attempt to find a empirical formula to depict this kind of relation.
Instead, we adopt simplified assumptions that sub-energetic SNe
generally produce more massive fallback disks, i.e., the logarithm
of the disk mass $\log (M_{d}/M_{\odot})$ is distributed uniformly
between $-5$ and $-1$ for stars of masses $8\leq M\leq12\,M_{\odot}$
or $M\geq 18\,M_{\odot}$, and between $-6$ to $-2$ in other cases.

The initial magnetic fields are chosen so that $\log B$ is
distributed normally with a mean of 12.4 and a standard deviation of
0.4.  The magnetic field decay parameter $M_{0}$ is set to be
$10^{-5}-10^{-4}\,M_{\odot}$, and the disk radius $R_{d}\sim
10^7-10^{8}$\,cm. The initial $P_{0}$ is chosen to be distributed
uniformly between 10 and 100\,ms.  In this work an MPSR is defined
as a radio pulsar with spin period less than 10\,ms, and we {\it
only} take the FDAI MPSRs (either in binaries or isolation) into
account. In the case of BMPSRs, if the companion star fills its
Roche lobe before the pulsar passes the death line, we take the
elapsed time before this moment as the upper limit for the lifetime
of the pulsar.

We divide the binary FDAI MPSRs into several categories according to
their formation channels and the properties of the companion stars
described as follows. The first case is that the pulsar is formed by
core collapse of the primary star, and has an MS binary companion.
Since the pulsar evolution can be strongly influenced by the stellar
winds from the companion, which are intensive for massive stars, we
only consider the MPSR-low-mass main sequence (MPSR-LMS) binaries,
in which the companion star has a mass less than $1.5\,M_{\odot}$.
Alternatively the pulsar is formed by core collapse of the
secondary, and already has an NS/BH binary companion. They are
called NS/BH-MPSRs binaries. Finally, a single MPSR could form
during either the first or the second SN that disrupts the binary.
%In this case, the velocity of the post-SN single MPSR would be same
%as the pre-SN orbital velocity of its progenitor, i.e. of tens of
%km\,s$^{-1}$, which is consistent with that of the disk MPSRs
%\citep{toscano99}. Moreover, SN kicks imparted on nascent NSs would
%aggravate disruptions.  If a binary is disrupted by a large kick
%($V_{\sigma}=265$\,km\,s$^{-1}$), the resultant single MPSR would
%In these casses, if the pulsar has a velocity greater than
%200\,km\,s$^{-1}$, we name it a ``fast" single MPSR; otherwise it is
%``slow".

To investigate the influence of the model parameters on the final
results, we design nine models by changing the values of the CE
parameter $\alpha_{CE}$\footnote{The CE parameter $\alpha_{CE}$ is
defined as the ratio of total binding energy of the envelope
$E_{bind,i}$, and the difference between the initial and final
orbital energies of the cores, i.e.,
$\alpha_{CE}=E_{bind,i}/(E_{orb,f}-E_{orb,i})$, where $E_{orb,f}$
and $E_{orb,i}$ are the final and initial binding energy of the
cores, respectively (see Hurley, Tout \& Pols 2002 for details).},
the field decay parameter $M_0$, the lower and upper limit of the
fallback disk mass $M_{d,low}$, $M_{d,upp}$ in case of sub-energetic
SNe, and the disk radius $R_{d}$ (listed in Table 2). The calculated
numbers and formation rates of various types of binary and single
MPSRs in our Galaxy are listed in Table 3, and briefly described as
follows.

In model A (regarded as our control model) we adopt
$\alpha_{CE}=1.0$, $M_{0}=10^{-5}\,M_{\odot}$,
$M_{d}=10^{-5}-0.1\,M_{\odot}$, and $R_{d}=10^{8}$\,cm. It is
predicted that in the Galaxy there are $\sim 200$ MPSR-LMS (PSR
J1903$+$0327 may be one of them) and ten times more NS/BH-MPSR
systems. Moreover, $\sim 10^{5}$ single MPSRs are produced. This
number is comparable with estimated from observations
\citep[e.g.][]{lorimer95}, suggesting that some fraction of the
single MPSRs might form from the fallback disk accretion rather the
traditional recycling channel in binary systems. Since their
formation rate is 100 times lower than normal pulsars, it is not
surprising that no MPSRs have been detected in young SN remnants.

In model B we increase the $\alpha_{CE}$ to 3.0, and find that
higher $\alpha_{CE}$ produces more MPSR-LMS binaries. This is
because binaries with a higher $\alpha_{CE}$ tend to survive during
the CE evolution, especially for low-mass, close binaries which are
more likely to be subject to coalescence during CE evolution.
%For a single MPSR, its progenitor binary with wider
%post-CE orbit would be more likely to be disrupted by the SN
%explosion.
For the same reason, lower $\alpha_{CE}$ (model C) leads to fewer
MPSRs \citep[see][for more detailed discussions]{liu06}. In model D
we adopt a slower magnetic field decay, resulting in a shorter
duration of the {\it accretor} phase, hence significantly reducing
the numbers of all three types of MPSRs. In model E we choose a
lower value of $R_{d}$ than in model A. As mentioned in \S 2, this
will cause longer final periods of the pulsars (compare the curves
of P1 and P3 in Fig.~1), and produce fewer MPSRs.  In model F the
lower limit of the fallback disk masses is increased by a factor of
3, but the numbers of MPSRs change little, since MPSRs are produced
only in the case of heavy disk accretion. This effect can also be
illustrated in model G: when the upper limit of disk masses become
half in model A, the numbers of MPSRs decrease by a factor of 10. To
constrain the possible ranges of MPSR numbers, models H and I
represent two extremely favorable and unfavorable cases for the
production of MPSRs, respectively.  It is found that the number of
MPSR-LMS binaries may range from less than one to nearly one
thousand in the Galaxy.

We plot the eccentricity vs. orbital period distributions of the
MPSR-LMS binaries for models A, B, and C in Figure 2.  The
distributions of the other three models are similar as that of model
A.  It is seen that in both models A and C most MPSR-LMS binaries
are in wide orbits ($P_{orb}>10^{3}$\,days) with eccentricities
$e>0.3$.  For model B, the binary population seems to be distributed
into two groups. One is similar as that in models A and C, the other
stretches across large ranges of $e$ ($\sim 0-1$) and $P_{orb}$
($\sim1-10^{3}$\,day) with a tendency of larger $e$ accompanied with
larger $P_{orb}$.

%% This section contains more display math examples, including unnumbered
%% equations (displaymath environment). The last paragraph includes some
%% examples of in-line math featuring a couple of the AASTeX symbol macros.

\section{Discussion and Conclusions}

%% The displaymath environment will produce the same sort of equation as
%% the equation environment, except that the equation will not be numbered
%% by LaTeX.
In this paper we suggest a born-fast formation channel for MPSRs in
the Galaxy. In this scenario, a newborn NS may experience
spinning-up and field decay phase via accretion from a fossil disk,
established as a result of fallback following a SN explosion. For
appropriate choices of initial parameters, the NS could become a low
field MPSR within $<10^3$ yr. In particular, this scenario can
naturally explain the properties (millisecond spin and eccentric
orbit) of the recently discovered BMPSR PSR J1903$+$0327, without
invoking stellar interactions within a GC or peculiar triple system.

Our population synthesis calculations also suggest a population of
single FDAI MPSRs in the Galaxy. The formation of single MPSRs has
not been well understood, especially how the pulsar has lost its
binary companion after mass transfer. The conventional explanation
is that the pulsar was recycled in a binary located in a GC, and
then ejected out due to stellar encounters, or the high-energy
radiation from the pulsar has evaporated its companion
\citep[e.g.][]{ruderman89}. The latter explanation may also require
a GC origin of the binary \citep{king05}. Although the fraction of
BMPSRs and LMXBs in GCs is much larger with respect to the fraction
in the Galactic field, it seems unlikely that all single MPSRs have
originated from GCs. The fallback disk accretion scenario might
provide a possible way for the formation of not only single MSPs
\citep{miller01} but also planetary systems around them
\citep{lin91}.

Obviously the formation scenario proposed in this work is subject to
many uncertainties, so the results in Table 3 should be regarded as
the optimistic cases. One of the biggest issues is to determine how
much fallback material would have enough angular momentum at the
time of collapse to allow the formation of a disk. Our population
synthesis calculations suggest that there may be tens to hundreds of
MPSR-LMS binaries lurking in our Galaxy. For PSR J1903$+$0327, a
relatively large amount of fallback disk mass ($\ga 0.1\,M_{\odot}$)
is required to produce the very short period ($<3$\,ms),  implying
that the progenitor star may have experienced an unusual SN-fallback
history. It is interesting to note that the large mass ($\sim
1.74\,M_{\odot}$) of the pulsar may suggest a massive ($\ga
18\,M_{\sun}$) progenitor star \citep{zhang08} and probably
intensive fallback during the SN explosion. To form a centrifugally
supported accretion disk around the NS, the specific angular
momentum of the fallback matter should be larger than that needed
for a circular orbit at the NS radius
$(GM_{NS}R_{NS})^{1/2}\sim1.4\times10^{16}$\,cm$^{2}$\,s$^{-1}$.
Modern stellar evolution calculations suggest, however, that the
majority of massive stars lose angular momentum via magnetic
processes very efficiently.  \citet{heger05} found that magnetic
torques decrease the final rotation rate of the collapsing iron core
by about a factor of $30-50$ compared with the nonmagnetic
counterparts. In their magnetic models there is a small amount of
matter ($\sim 0.5\,M_{\odot}$ for a $15\,M_{\odot}$ star and
$\sim1.0\,M_{\odot}$ for a $25\,M_{\odot}$ star) that has specific
angular momentum greater than $10^{16}$\,cm$^{2}$\,s$^{-1}$ in the
envelope of the pre-SN progenitor star.  The realistic amount of
fallback matter would be significantly less than these values
because material with high specific angular momentum materials lies
in the envelope of the progenitor star rather than in its inner
core. Solutions of the problem of having enough rotation may involve
tidal interactions in close binaries. If either star in the binary
fills its Roche lobe, the spin-orbit coupling tidal torques would
corotate the spins of two components with the orbital rotation, so
that the progenitor star of the NS could be spun-up enough to allow
the formation of the fallback disk. If this is true the realistic
number of FDAI MPSRs should be less than our calculation, because
not all the progenitors of the FDAI MPSRs have experienced the close
binary evolution phases.

%Modern stellar
%evolution models predict that pulsars derived from more massive
%stars rotate faster \citep{heger05}, so it is possible that PSR
%J1903$+$0327 was born spinning fast (i. e. $<10$\,ms) considering
%its unusually high mass.

Other uncertainties include accretion from the SN fallback disk and
magnetic field decay induced by accretion. The rate at which
material falls back on the accretion disk depends on both the
density profile of the star and on its angular velocity profile.
Numerical calculations have been carried for a few mass and angular
momentum configurations of massive stars \citep[cf.][and references
therein]{woosley06}, which generally suggest highly super-Eddington
accretion with a short time. Still lack is a comprehensive picture
of the conditions for fallback disk formation after SN explosions of
massive stars. Highly super-Eddington accretion in a short period of
$\sim100$\,yr in a radiation-trapped regime plays a key role in the
formation of FDAI MPSRs.  This distinguishes our model from the
standard type of accretion expected in the recycling scenario, which
lasts for $\sim10^{8}-10^{10}$\,yr. In the standard pulsar recycling
theory, the so called ``spin-up line" in the $B-P$ diagram shows the
minimum period to which the NS can be spun-up in Eddington-limited
accretion. This line is defined by the equilibrium period assuming
the spin-up proceeding at the Eddington accretion rate.  Although
the pulsars in our scenario have experienced the highly
super-Eddington accretion, since the mass accretion rate declines
rapidly, the equilibrium period is never attained, unlike in the
case of accreting NS in LMXBs. Thus our model does not necessarily
predict any MPSRs sitting above the spin-up line.

Nor is known how accretion-induced field decay occurs in NSs. One
possible way is via ohmic dissipation of the accreting NSs crustal
currents, due to the heating of the crust which in turn increases
the resistance in the crust \citep{romani90,geppert94,konar97}.
Alternative scenarios consist of screening or burying the magnetic
field with the accreted material \citep{bisnov74,taam86,cumming01},
or outward moving vortices in the superfluid and superconducting
core pushing magnetic fluxoids into the crust during pulsar's
spin-down \citep{srinivasan90,konar99}. In this work we require a
fast field decay during the hyper-accretion phase (within $\la
10^3-10^4$ yr). Screening or burying the magnetic field with the
accreted material may be more appropriate for this scenario.

Finally we move to the uncertainties is the CE evolution. We have
used a constant CE parameter $\alpha_{CE}$ to compute the orbital
evolution during the spiral-in process, which is, however, very
likely to change with the properties and evolutionary state of stars
\citep{iben93,taam00}. From Fig.~2 we find that model B with
$\alpha_{CE}=3$ seems to be preferred for the formation of PSR
J1903. This value is be compatible with other estimates by, e.g.,
\citet{heuvel94}, \citet{portegies98} and \citet{kalogera99}, but in
contradiction with \citet{taam96}, \citet{sandquist98, sandquist00},
and \citet{oshau08}. Here the controversial issue is that there are
no strict criteria for defining binding energy of stellar envelopes
and there is no clear understanding whether sources other than
gravitational energy may contribute to unbinding common envelopes
\citep{han95}.

A distinct feature of the born-fast scenario is that the companion
star of PSR J1903$+$0327 is predicted to be a ``young" ($\la
1-2\times 10^9$ yr, i.e. within the characteristic age of the
pulsar) MS star. Future optical, IR, and radio observations could
present strong constraints on the nature of the optical counterpart,
and thus verify or falsify the born-fast scenario.

\acknowledgments

We thank X. -P. Zheng, H. -L. Dai and Z. -Y. Zuo for helpful
discussions, and J. R. Hurley for providing the BSE code.  It is
grateful to the anonymous referee for his/her insightful comments
that help improving the original manuscript.  This work was
supported by the Natural Science Foundation of China under grant
numbers 10573010 and and 10221001. XWL thanks Huazhong Normal
University for a postdoctoral fellowship.

\clearpage

%% the file, uncomment the \plotone, \plottwo, and
%% If you need a layout that cannot be achieved with \plotone or
%% \plottwo, you can invoke the graphicx package directly with the
%% \includegraphics command or use \plotfiddle. For more information,
%% please see the tutorial on "Using Electronic Art with AASTeX" in the
%% documentation section at the AASTeX Web site,
%% http://www.journals.uchicago.edu/AAS/AASTeX.
%%
%% The examples below also include sample markup for submission of
%% supplemental electronic materials. As always, be sure to check
%% the instructions to authors for the journal you are submitting to
%% for specific submissions guidelines as they vary from

%% This example uses \plotone to include an EPS file scaled to
%% 80% of its natural size with \epsscale. Its caption
%% has been written to indicate that additional figure parts will be
%% available in the electronic journal.

\begin{figure}
\epsscale{.80} \plotone{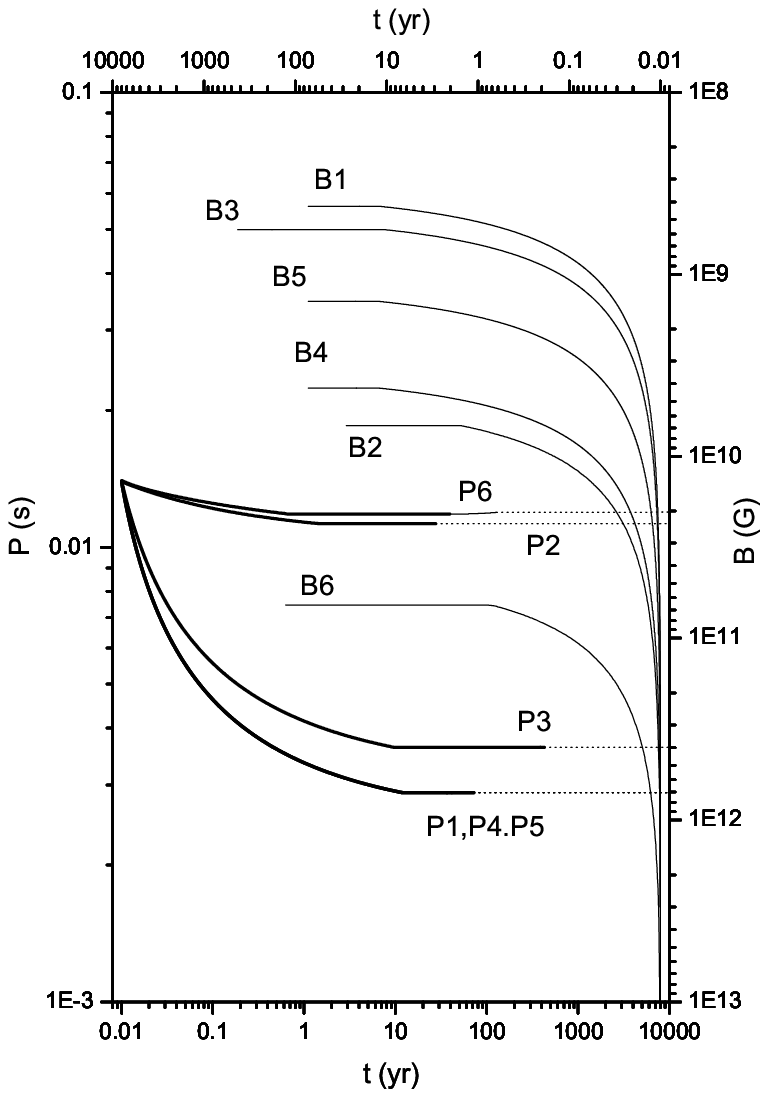} \caption{Period (bottom-left axes)
and magnetic field (top-right axes) evolution of an NS with initial
parameters list in Table 1.  The curves labeled P1 - P6 and B1 - B6
represent the evolutions of period and magnetic field, respectively.
Numbers are suffixed to indicate the corresponding models.  In the
$P-t$ plot, the thick solid lines, solid lines, and dotted lines
represent the {\it accretor}, the {\it propeller}, and the {\it
radio pulsar} phases, respectively.  The magnetic field evolution is
plotted until the {\it radio pulsar} phase begins. \label{fig1}}
\end{figure}

\clearpage

\begin{figure}
%\epsscale{.80}
\plotone{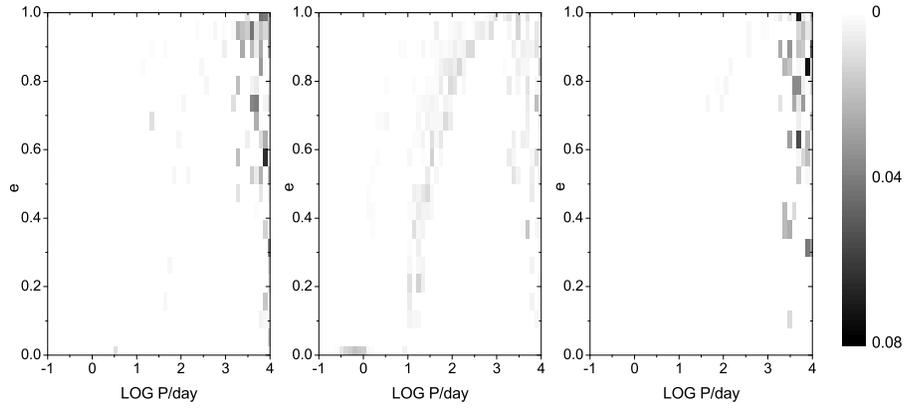} \caption{Orbital period and eccentricity
probability distribution of the MPSR-LMS binaries for model A, B,
and C, from left to right, respectively. The distributions for the
other six models (D-G, H, and I) are very similar to those for
models A, B, and C.\label{fig2}}
\end{figure}

%\clearpage

%% Here we use \plottwo to present two versions of the same figure,
%% one in black and white for print the other in RGB color
%% for online presentation. Note that the caption indicates
%% that a color version of the figure will be available online.
%%

%% This figure uses \includegraphics to scale and rotate the still frame

%% If you are not including electonic art with your submission, you may
%% mark up your captions using the \figcaption command. See the
%% No more than seven \figcaption commands are allowed per page,
%% so if you have more than seven captions, insert a \clearpage

%% Tables should be submitted one per page, so put a \clearpage before

%% Two options are available to the author for producing tables:  the
%% deluxetable environment provided by the AASTeX package or the LaTeX
%% table environment.  Use of deluxetable is preferred.
%%

%% Three table samples follow, two marked up in the deluxetable environment,
%% one marked up as a LaTeX table.

%% In this first example, note that the \tabletypesize{}
%% command has been used to reduce the font size of the table.
%% We also use the \rotate command to rotate the table to
%% landscape orientation since it is very wide even at the
%% Note also that the \label command needs to be placed

%% This table also includes a table comment indicating that the full
%% version will be available in machine-readable format in the electronic

\clearpage

%% If you use the table environment, please indicate horizontal rules using
%% Do not put multiple tabular environments within a single table.
%% The optional \label should appear inside the \caption command.

\begin{table}
\begin{center}
\caption{Initial model parameters for the NS-disk
systems.\label{tbl-2}}
\begin{tabular}{cccccc}
\tableline\tableline Model No. & $M_{d}$\,(M$_{\odot}$) &
$R_{d,10}$ & $P_{0}$\,(ms) &
$B_{0}$\,($10^{12}$G) & $M_{0}$\,($10^{-5}$M$_{\odot}$) \\
\tableline
1 & 0.28 & 0.1 & 14 & 3 & 1 \\
2 & 0.02 & 0.1 & 14 & 3 & 1 \\
3 & 0.28 & 0.01 & 14 & 3 & 1 \\
4 & 0.28 & 0.1 & 14 & 3 & 10 \\
5 & 0.28 & 0.1 & 14 & 10 & 1 \\
6 & 0.01 & 0.01 & 14 & 10 & 1 \\

\tableline
\end{tabular}
%% Any table notes must follow the \end{tabular} command.

\end{center}
\end{table}

\begin{table}
\begin{center}
\caption{Model parameters for binary population
synthesis.\label{tbl-2}}
\begin{tabular}{cccccc}
\tableline\tableline Model & $\alpha_{CE}$ &  $M_{0}$
($10^{-5}$\,M$_{\odot}$) & $R_{d,8}$ & $M_{d,low}$ ($10^{-5}$\,M$_{\odot}$) & $M_{d,upp}$ ($0.1$\,M$_{\odot}$)\\
\tableline
A & 1 &  1 & 1 & 1 & 1\\
B & 3 &  1 & 1 & 1 & 1 \\
C & 0.5 &  1 & 1 & 1 & 1\\
D & 1 &  10 & 1 & 1 & 1\\
E & 1 &  1 & 0.1 & 1 & 1\\
F & 1 &  1 & 1 & 3 & 1\\
G & 1 & 1 & 1 & 1 & 0.5\\
H & 3 & 1 & 1 & 3 & 1\\
I & 0.5 & 10 & 0.1 & 1 & 0.5\\
 \tableline
\end{tabular}
%% Any table notes must follow the \end{tabular} command.
%\tablenotetext{a}{The lower limit of the fallback disk mass of the
%weak SN explosions, i.e. for initial stellar masses of $8\leq
%M\leq12$\,M$_{\odot}$ or $M\geq 18$\,M$_{\odot}$}
\end{center}
\end{table}

\begin{table}
\begin{center}
\caption{Predicted numbers and formation rates in our Galaxy of
various types of FDAI MPSR systems.\label{tbl-2}}
\begin{tabular}{ccccc}
\tableline\tableline Model & item & MPSR-LMS
& NS/BH-MPSR & Single MPSR \\
\tableline
A & number & 203 & $3.3\times10^{3}$ &  $1.5\times10^{5}$\\
  & rate   & $6.0\times 10^{-7}$  & $3.1\times 10^{-6}$ & $3.1\times 10^{-4}$\\
\tableline

B & number & 751 & $3.7\times10^{3}$ & $1.6\times10^{5}$\\
  & rate   & $1.3\times 10^{-6}$  & $4.0\times 10^{-6}$ & $3.2\times 10^{-4}$ \\
\tableline

C & number & 192 & $3.1\times10^{3}$ & $1.5\times10^{5}$\\\
  & rate   & $6.5\times 10^{-7}$  & $3.0\times 10^{-6}$ & $3.1\times 10^{-4}$\\
\tableline

D & number & 8 & $1.4\times10^{2}$ & $1.2\times10^{3}$\\
  & rate   & $6.2\times 10^{-7}$  & $3.3\times 10^{-6}$ & $2.7\times 10^{-4}$ \\
\tableline

E & number & 83 & $1.3\times10^{3}$ & $4.9\times10^{4}$ \\
  & rate   & $3.9\times 10^{-7}$  & $2.0\times 10^{-6}$ & $1.9\times 10^{-4}$ \\
\tableline

F & number & 232 & $3.7\times10^{3}$ & $1.7\times10^{5}$ \\
  & rate   & $6.9\times 10^{-7}$  & $3.5\times 10^{-6}$ & $3.4\times 10^{-4}$ \\
\tableline

G & number & 28 & $4.5\times10^{2}$ & $1.2\times10^{4}$ \\
  & rate   & $3.5\times 10^{-7}$  & $1.7\times 10^{-6}$ & $2.0\times 10^{-4}$ \\
\tableline

H & number & 856 & $4.2\times10^{3}$ & $1.8\times10^{5}$ \\
  & rate   & $1.5\times 10^{-6}$  & $4.5\times 10^{-6}$ & $3.6\times 10^{-4}$ \\
\tableline

I & number & 0.3 & 7 & 38 \\
  & rate   & $9.9\times 10^{-8}$  & $4.6\times 10^{-7}$ & $3.9\times 10^{-5}$ \\
\tableline

\end{tabular}
%% Any table notes must follow the \end{tabular} command.

\end{center}
\end{table}

%% If the table is more than one page long, the width of the table can vary
%% from page to page when the default \tablewidth is used, as below.  The
%% individual table widths for each page will be written to the log file; a
%% maximum tablewidth for the table can be computed from these values.
%% The \tablewidth argument can then be reset and the file reprocessed, so
%% that the table is of uniform width throughout. Try getting the widths
%% from the log file and changing the \tablewidth parameter to see how
%% adjusting this value affects table formatting.

%% The \dataset{} macro has also been applied to a few of the objects to
%% show how many observations can be tagged in a table.

\clearpage

%% Tables may also be prepared as separate files. See the accompanying
%% sample file table.tex for an example of an external table file.
%% To include an external file in your main document, use the \input
%% command. Uncomment the line below to include table.tex in this
%% sample file. (Note that you will need to comment out the \documentclass,
%% \begin{document}, and \end{document} commands from table.tex if you want
%% to include it in this document.)


%% The following command ends your manuscript. LaTeX will ignore any text

\end{document}